\documentstyle[onecolumn,epsf,mnras_cite]{mn}
\title{The correlation of peaks in the microwave background}
\author[Alan F. Heavens and Ravi K. Sheth]
{Alan F. Heavens$^1$ and Ravi K. Sheth$^2$\\ 
$^1$Institute for Astronomy, University of Edinburgh, Blackford Hill,
Edinburgh EH9 3HJ, U.K.\\ $^2$MPI f\"ur Astrophysik, Karl
Schwarzschild Str. 1, D85740 Garching b. M\"unchen, Germany }

\newcommand{\be}{\begin{equation}}
\newcommand{\ee}{\end{equation}}
\newcommand{\ba}{\begin{eqnarray}}
\newcommand{\ea}{\end{eqnarray}}

\newcommand{\nn}{\nonumber\\}

\newcommand{\bk}{{\bf k}}

\newcommand{\br}{{\bf r}}

\newcommand{\bv}{{\bf v}}
\newcommand{\bx}{{\bf x}}
\newcommand{\dd}{{\partial}}

\def\gs{\mathrel{\raise1.16pt\hbox{$>$}\kern-7.0pt 
\lower3.06pt\hbox{{$\scriptstyle \sim$}}}}         
\def\ls{\mathrel{\raise1.16pt\hbox{$<$}\kern-7.0pt 
\lower3.06pt\hbox{{$\scriptstyle \sim$}}}}         

\begin{document}

\maketitle

\begin{abstract}
We present accurate predictions of the correlation function of
hotspots in the microwave background radiation for gaussian theories
such as those predicted in most inflation models.  The correlation
function of peaks above a certain threshold depends only on the
threshold and the power spectrum of temperature fluctuations.  Since
there are both potentially observable quantities in a microwave
background map, there are no adjustable parameters in the predictions.
These correlations should therefore provide a powerful test of the
gaussian hypothesis, and provide a useful discriminant between
inflation and topological defect models such as the cosmic string
model.  The correlations have a number of oscillatory features, which
should be detectable at high signal-to-noise with future satellite
experiments such as MAP and Planck.
\end{abstract}

\begin{keywords}
cosmic background radiation - cosmology; theory - early Universe -
large-scale structure of Universe.
\end{keywords}

\section{Introduction}

The microwave background radiation is principally a relic of the early
Universe, containing information on the fluctuations present at a
redshift of about 1000.  Detailed knowledge of the fluctuations is
extremely valuable as a tool for distinguishing structure formation
theories.  The main alternative to the microwave background for this
purpose is the distribution of galaxies at the present day, or at
moderate redshifts.  The microwave background has two great advantages
over galaxy redshift surveys.  Firstly, one needs to make no
assumptions about the relationship between galaxy clustering and mass
clustering (the issue of bias), and secondly, the microwave background
probes the Universe when the fluctuations were extremely small, so
linear perturbation theory can describe the predicted fluctuations to
high accuracy.  Against these advantages must be offset the problems
of foreground contamination and the technical difficulties of
producing a high signal-to-noise, high-resolution map of large areas
of sky.  Future satellite experiments planned by ESA and NASA
(\pcite{PhaseA}, \pcite{Jungman}) will produce such maps, and these
will allow the spectrum of temperature fluctuations to be measured
with high accuracy.  The primary goal of such experiments will be to
determine a number of cosmological parameters, such as the density
parameter $\Omega_0$, the Hubble constant, and the mix of species of
dark matter, all of which, in inflationary theories, influence the
detailed shape of the temperature power spectrum.  In such theories,
the temperature map is a random gaussian field, whose properties are
determined by the power spectrum (or correlation function) alone.  It
is vitally important to test whether the map is indeed gaussian, as if
it is not, the proposed parameter estimation method
(e.g. \pcite{Jungman}) is invalid.  In addition, a non-gaussian map of
the primary temperature anisotropies would indicate a quite different
structure formation scenario from the standard inflation model.

There are several ways to test the gaussian hypothesis, such as the
3-point function (e.g. \pcite{Hinshaw94}, \pcite{FRS93}, \pcite{LS93},
\pcite{GLMM94}), the genus and Euler-Poincar\'e statistic
(\pcite{Coles88}, \pcite{Gott90}, \pcite{Luo94b}, \pcite{Smoot94}),
the bispectrum (\pcite{Luo94}, \pcite{Hea98}, \pcite{Ferreira98}),
studies of tensor modes in the CMB \cite{CCT94}, and peak statistics
(\pcite{BE87}, \pcite{Kogut95}, \pcite{Kogut96}).  For all of these it
is possible in principle to make predictions for gaussian fields, and
each method has its advantages.  The bispectrum is attractive because
the error analysis can be done without recourse to numerical
simulation, and can treat the twin problems of missing sky regions and
correlated noise \cite{Hea98}, albeit at large computational expense.
In this paper we explore an alternative: the correlation function of
peaks in the microwave background.  This has the advantage of being
readily calculable, and also we find that the predicted correlation
functions have oscillatory features which are not obviously present in
the temperature autocorrelation function.  In much the same way as the
oscillations in the power spectrum allow high precision determinations
of cosmological parameters, the peak correlation function oscillations
should allow a high precision test of the gaussian hypothesis.  In the
latter case, the correlation function of peaks above any given
threshold is completely determined once the power spectrum is known;
there are no free parameters.

In this paper, we calculate the 2-point correlation function of peaks
in a 2D gaussian field, for any given power spectrum.  We make no
approximations other than to compute the correlation function for
small angular separations so we assume the sky is flat and translate
the sky power spectrum to an analogous 2D spectrum.  We follow the
peak statistics methods developed initially by \scite{PH85} and
\scite{BBKS}.

The layout of the paper is as follows.  In section 2 we describe the
method, and in section 3 present the results and compare with an
approximate correlation function method introduced by \scite{BBKS}.
In section 4 we discuss the results and the possibilities for Planck
and MAP.

\section{Method}

In this section, we compute the two-point correlation function of
local maxima in 2D gaussian random fields.  For small separations,
this calculation should be accurately applicable to the clustering of
peaks in the microwave sky (see \scite{BE87} for further discussion of
this point).  The calculation is laid out as follows: after some
definitions, we compute the $12 \times 12$ covariance matrix for the
12 variables which are required to specify the two peaks.  We then
invert this matrix to obtain the multivariate gaussian distribution
for the variables, then apply the peak constraint and integrate over
the remaining variables to obtain the correlation function.

\subsection{Peaks in 2D}

We define the temperature fluctuation by $\delta(\bx) \equiv
T(\bx)/\bar T -1$, and its 2D Fourier transform by $\delta_\bk \equiv
\int d^2\bk\, \delta(\bx) \exp(i\bk\cdot\bx)$.   The fluctuations are
specified entirely by the power spectrum, $P(k)$, defined by
\be
\langle \delta_\bk \delta_{\bk'}\rangle = (2\pi)^2 P(k) \delta^D(\bk+\bk')
\label{Pk}
\ee
where angle brackets indicate ensemble averages, and $\delta^D$ is a
2D Dirac delta function.  We will require moments of the power
spectrum, defining
\be
\sigma_j^2 \equiv {1\over (2\pi)^2} \int d^2\bk \,P(k) k^{2j},
\ee
and spectral parameters 
\be
\gamma \equiv \sigma_1^2/(\sigma_0
\sigma_2)  \qquad\theta_* \equiv \sqrt{2}{\sigma_1\over \sigma_2}.
\label{Parameters}
\ee 
At peaks, the 2D gradient $\delta_i \equiv \dd \delta/\dd x_i$
vanishes, and the eigenvalues of the second derivative matrix
$\delta_{ij} \equiv \dd^2 \delta/\dd x_i \dd x_j$ are all negative.
Since $\delta_{ij}$ is symmetric, we are concerned with 6 independent
numbers $\bv =
\{\delta,\delta_x,\delta_y,\delta_{xx},\delta_{xy},\delta_{yy}\}$ to
specify a local maximum.  For a gaussian field, the probability
density function for the 12 variables $\bv \equiv
\{\bv^{(1)},\bv^{(2)}\}$, evaluated at two points separated by a
distance $\br$, is 
\be 
p(\bv_1,\bv_2) = {1\over (2\pi)^6 ||M||^{1/2}}
\exp\left(-{1\over 2} v_i M^{-1}_{ij} v_j\right).  
\ee 
The covariance
matrix is $M_{ij} \equiv \langle (v_i - \bar v_i)(v_j - \bar
v_j)\rangle$, overlines indicate ensemble averages, and the summation
convention is assumed.  Note that
$\bar v_i=0$ in our case, and $M^{-1}_{ij}$ is the $i,j$ component of
the inverse of $M$.

The number density of peaks is a sum of 2D Dirac delta functions,
centred on peaks $\bx_{pk,q}$:
\be
n_{pk} = \sum_q \delta^D(\bx - \bx_{pk,q})
\label{n}
\ee
In the vicinity of a peak, we may expand
\be
\delta(\bx) \simeq \delta(\bx_{pk}) + {1\over 2}\sum_{i,j} {\dd^2
\delta\over \dd x_i \dd x_j} (\bx - \bx_{pk})_i (\bx - \bx_{pk})_j
\ee
so
\be
(\bx - \bx_{pk})_i  \simeq \left({\dd^2
\delta\over \dd x_i \dd x_j}\right)^{-1} {\dd \delta\over \dd x_j}
\ee
and the Dirac delta function in (\ref{n}) is 
\be
\delta^D(\bx - \bx_{pk}) \simeq |\det \delta_{ij}| \delta^D(\delta_i).
\ee
The mean number density is
\ba
\langle n_{pk}(\bx)\rangle& =& \langle | \det \delta_{ij}|
\delta^D(\delta_i) \rangle\nn
& = & \int d\delta \,d \delta_{xx}\, d \delta_{xy}\, d \delta_{yy}\,  | \det
\delta_{ij}|\, p(\delta, \delta_i=0, \delta_{ij}) 
\ea
and the range of integration is set by the requirement that the
eigenvalues of $-\delta_{ij}$ are all positive.  These eigenvalues are
\be
\lambda = {1\over 2}\left[-(\delta_{xx}+\delta_{yy}) \pm
\sqrt{(\delta_{xx}-\delta_{yy})^2 +4 \delta_{xy}^2}\right].
\ee
If we introduce the notation
\ba
\nu & \equiv & {\delta\over \sigma_0}\nn
X & \equiv & -{(\delta_{xx}+\delta_{yy})\over \sigma_2}\nn
Y &  \equiv & {(\delta_{xx}-\delta_{yy})\over \sigma_2}\nn
Z & \equiv & {2\delta_{xy}\over \sigma_2}
\label{Vars}
\ea
then the conditions for a maximum are $X>0$ and $Y^2+Z^2<X^2$.  The
determinant is
\be
\det \delta_{ij} = {X^2-Y^2-Z^2\over 4\sigma_2^2}.
\ee
To compute the mean number density, we need the covariance matrix for
the 6 variables specifying the field and its first and second
derivatives.  The non-zero components are $\langle \nu^2 \rangle =
\langle X^2\rangle = 1$;   $\langle \nu X \rangle = \gamma$; 
$\langle Y^2 \rangle =  \langle Z^2 \rangle
= 1/2$; $\langle \delta_x^2 \rangle = \langle \delta_y^2 \rangle =
\sigma_1^2/2$.   The matrix can readily be inverted, and the integrations
over $Y$, $Z$ and $X$ performed.    The expression for the
differential number density as a function of peak height, $n_{pk}(\nu)$ is
given in \scite{BE87}, equation (A1.9):
\be
n_{pk}(\nu)={1\over (2\pi)^{3/2} \theta_*^2}\exp(-\nu^2/2)\,G(\gamma,\gamma\nu)
\ee
where
\ba
G(\gamma,x_*)&\equiv& (x_*^2-\gamma^2)\left\{1-{1\over 2}{\rm
erfc}\left[ {x_*\over \sqrt{2(1-\gamma^2)}} \right]\right\} +
x_*(1-\gamma^2) {\exp\left[-x_*^2\over 2(1-\gamma^2)\right]\over
\sqrt{2\pi(1-\gamma^2)}} + \nn
&&{\exp\left[-x_*^2\over 3-2\gamma^2\right]\over
\sqrt{3-2\gamma^2}}\left\{1-{1\over 2}{\rm erfc}\left[{x_*\over \sqrt{2(1-\gamma^2)(3-2\gamma^2)}}\right]\right\}
\ea
The correlation function is obtained similarly.  We find the
covariance matrix for the 12 variables, invert it to find their
probability distribution, and integrate subject to the
constraints. The cross-correlation function of peaks of height $\nu_1$
and $\nu_2$ is then given by
\ba
1+\xi(r| \nu_1, \nu_2) &= & {1\over 4\theta_*^4  n_{pk}(\nu_1) n_{pk}(\nu_2)}
\int_{X_1=0}^\infty\int_{X_2=0}^\infty
\int_{Y_1=-X_1}^{X_1}\int_{Y_2=-X_2}^{X_2}
\int_{Z_1=-\sqrt{X_1^2-Y_1^2}}^{\sqrt{X_1^2-Y_1^2}}\int_{Z_2=-\sqrt{X_2^2-Y_2^2}}^{\sqrt{X_2^2-Y_2^2}}
dX_1 dX_2 dY_1 dY_2 dZ_1 dZ_2 \times \nn
& & \left(X_1^2-Y_1^2-Z_1^2\right)\left(X_2^2-Y_2^2-Z_2^2\right)
p(\nu_1,X_1,Y_1,Z_1,\delta^{(1)}_{x}=0,
\delta^{(1)}_{y}=0, \nu_2,X_2,Y_2,Z_2,\delta^{(2)}_{x}=0,
\delta^{(2)}_{y}=0).
\label{Xsi}
\ea
The correlation function for peaks above a certain threshold $\nu$ is
obtained by adding two further integrations over $\nu_1$ and $\nu_2$,
and replacing the differential number densities $n_{pk}(\nu)$ in the
denominator of (\ref{Xsi}) by numerically-evaluated integrals
$n_{pk}(>\nu)$.  The calculation of the covariance matrix and the
joint probability distribution is rather technical, and so appears in
an appendix.  

\section{Results}

In the flat-sky approximation, the power spectrum $P(k)$ which we
require is related simply to the conventional power spectrum of
spherical harmonic coefficients $C_\ell \equiv \langle |a_{\ell
m}|^2\rangle$, where $\delta T(\theta,\phi) \equiv \sum_{\ell m}
a_{\ell m} Y_\ell^m(\theta,\phi)$
\cite{BE87}:
\be
P(k) = C_\ell \quad {\rm with\ }k\simeq \ell
\ee
We run CMBFAST \cite{SZ96} to generate the power spectrum $C_\ell$,
and interpolate to obtain a continuous power spectrum $P(k)$.  We
model the beam with a gaussian of FWHM $b$, so multiply the power
spectrum by a gaussian $\exp\left[-\sigma^2 \ell(\ell+1)\right]$, with $\sigma=
b/\sqrt{8\ln2}$.    For peaks above a threshold, an 8D numerical integration  
is performed (although one could be done analytically), and, despite
the integrand consisting of about 3000 lines of code, each point in the correlation
function can be computed accurately in tens of seconds on a fast workstation.

\begin{figure}
\centering
\begin{picture}(200,200)
\includegraphics{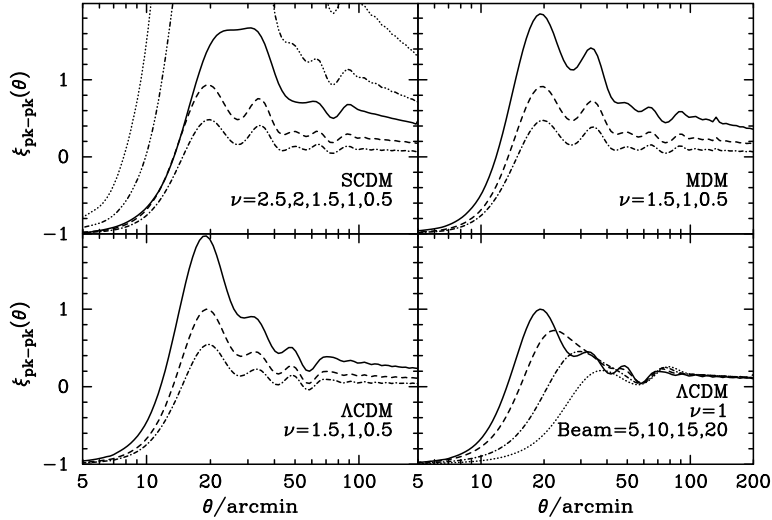}
\end{picture}
\caption[]{\label{CDMpanels} The correlation function of peaks in the
microwave background, for a cold dark matter (CDM) model (top left;
baryon density $\Omega_B=0.05$, $\Omega_{CDM}=0.95$), a mixed dark
matter model containing hot dark matter (top right; $\Omega_B=0.05$,
$\Omega_{CDM}=0.8$, $\Omega_{HDM}=0.15$), and a flat cosmological constant
model (bottom left; $\Omega_B=0.05$, $\Omega_{CDM}=0.15$,
$\Omega_\Lambda=0.8$).  A Hubble constant $H_0=60$ km s$^{-1}$
Mpc$^{-1}$ is assumed and the beam is 5 arcminutes FWHM.  At bottom
right, we show the effect on the cosmological constant model of
altering the beam size.  5, 10, 15 and 20 arcminute beams are shown.}
\end{figure}

In Fig. \ref{CDMpanels}, we show the peak-peak correlation
functions for a variety of models, for a 5 arcminute beam and peaks
above various thresholds.  The striking feature of
these graphs is the wealth of structure in the correlation functions.
The first peak is not due to the beam, rather due to the sharp
turndown in the power spectrum itself at around $\ell=1500$, due to
the thickness of the last scattering surface and/or Silk damping.  There
may be a correspondence in the peak positions with analogous peaks in
the power spectrum, but it could also be ringing due to the abrupt
cutoff.  Note in particular that the
temperature autocorrelation function (the Fourier transform of the
power spectrum) is very smooth for these separations, without any
oscillatory features.
%
%
%
%
%
%
We also show in Fig. \ref{CDMpanels} the effect of beam-smearing.
Correlation functions for beams of 5, 10, 15 and 20 arcminute are
shown.  The oscillations are more numerous and pronounced for smaller
beam sizes, so it is advantageous to work at high resolution.

\begin{figure}
\centering
\begin{picture}(200,200)
 \includegraphics{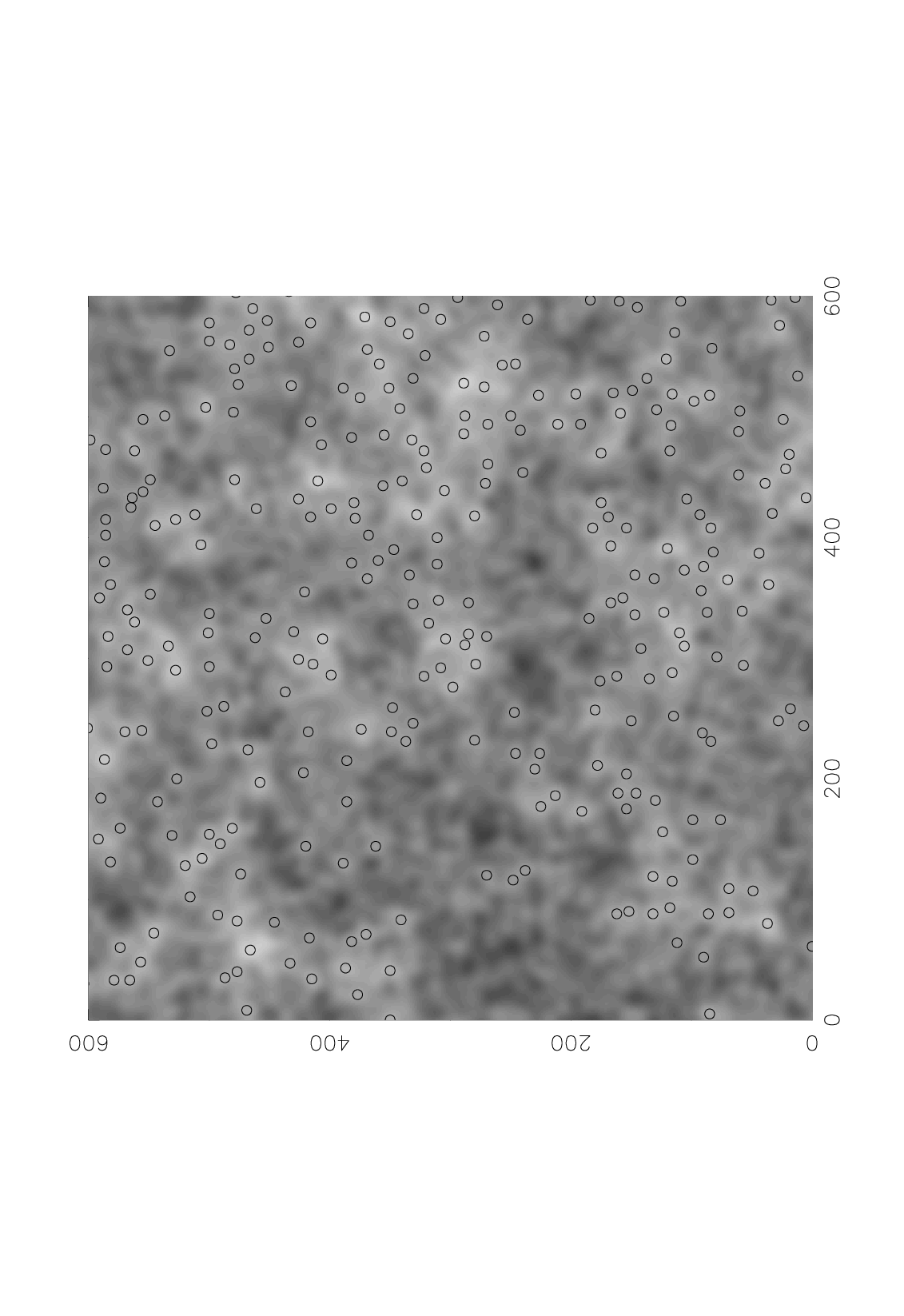}
\end{picture}
\caption[]{\label{Picture} A $10^\circ \times 10^\circ$ patch of a
mixed dark matter simulated sky, at 5 arcminute resolution
characteristic of the highest frequency Planck channels.  Peaks above
$1\sigma$ are marked with circles.}
\end{figure}

\begin{figure}
\centering
\begin{picture}(200,200)
 \includegraphics{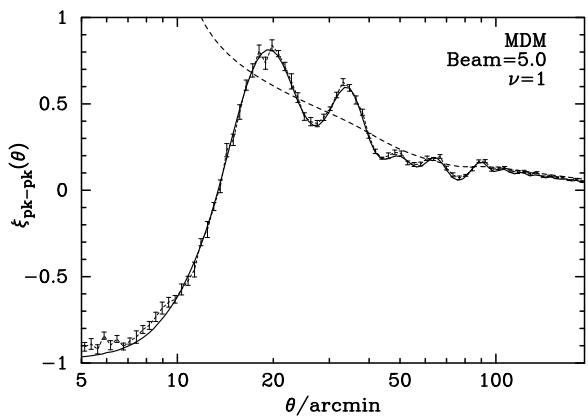}
\end{picture}
\caption[]{\label{MDMNum}  The $\nu=1$ threshold from the mixed dark
matter model in Fig. \ref{CDMpanels},
with the correlation function from simulated gaussian fields
superimposed.  Also shown is the approximate correlation function
computed by \scite{BE87} by ignoring derivatives of the temperature
autocorrelation function.}
\end{figure}

Later in this paper we will want to compute the correlations of
estimates of the peak-peak correlation function.   This is difficult
analytically, so we have generated 
sky realisations, of $3072 \times
3072$ pixels, each 1 arcminute across, located local maxima, and
computed the correlation function (see Fig. \ref{Picture}).  Edge
effects are irrelevant as we wrap the box round for pair counting.
Fig. \ref{MDMNum} shows the average of 10 such realisations, with the
error on the mean plotted for each bin.  The beam is 5 arcminutes and
the spectrum is from a mixed dark matter model, which we have apodized
with a cosine bell to remove power at long wavelengths.  The apodizing affects
$k<30$, and is introduced to avoid discreteness errors arising at
low-$k$ from our continuum approximation used to evaluate the moments
of the power spectrum and the $\lambda_{\alpha\beta\gamma}$, defined
in the Appendix.  Note the good agreement with the theoretical curve.

\begin{figure}
\centering
\begin{picture}(200,200)
 \includegraphics{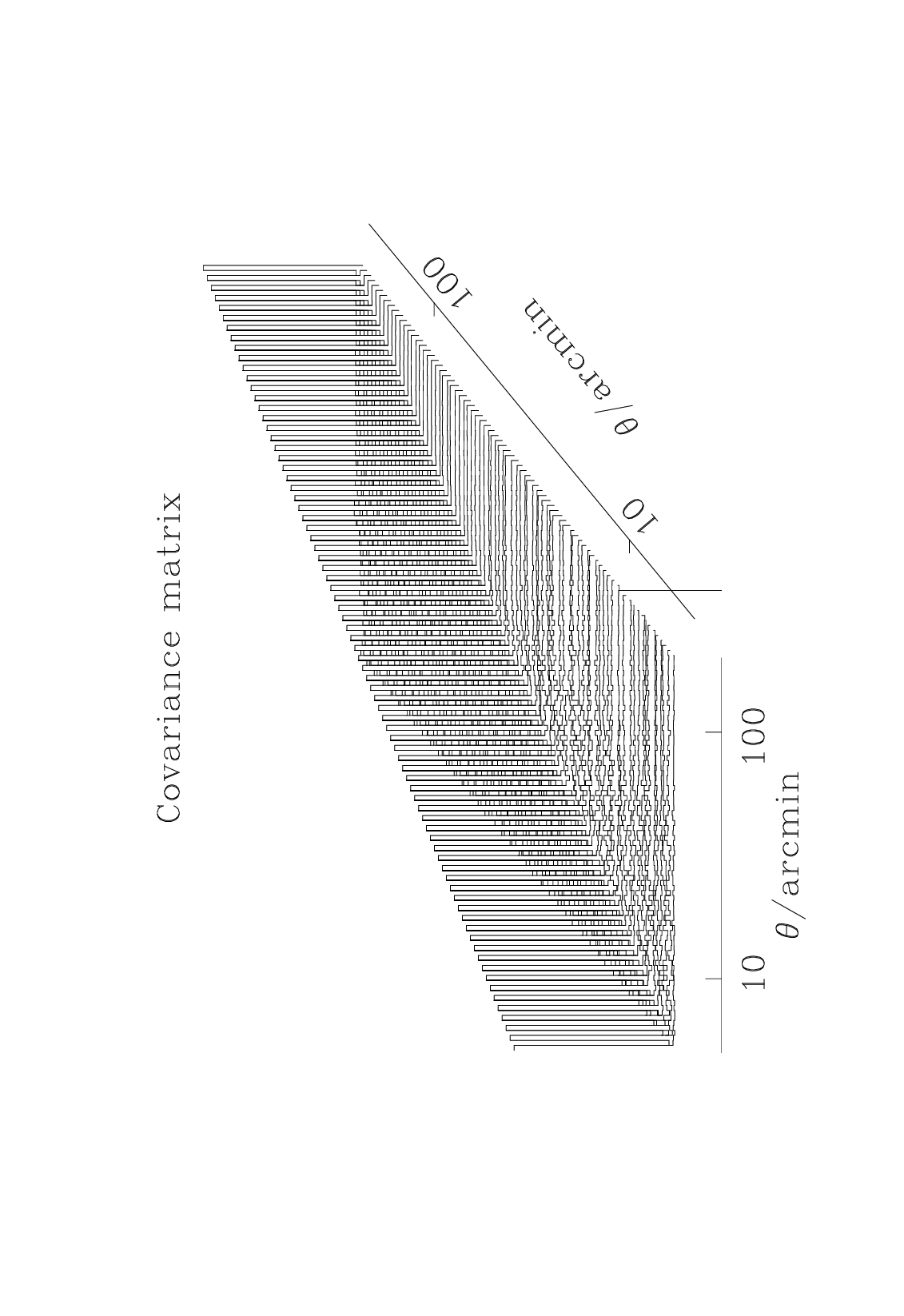}
\end{picture}
\caption[]{\label{Covariance} Correlation coefficients for estimators of
the peak-peak correlation function of peaks above $1\sigma$ in the MDM
simulation shown in Fig. \ref{Picture}.}
\end{figure}

This numerical approach allows us to compute the correlation
coefficients of the estimators $\hat\xi$ of the peak-peak correlation
function.  In Fig. \ref{Covariance} we show, from 100 realisations of a
mixed dark matter model, $\langle (\hat\xi_i -
\bar\xi_i)(\hat\xi_j - \bar\xi_j)\rangle/(\sigma_i\sigma_j)$ where
$\bar \xi_i$ is the mean value and $\sigma_i$ the variance of the
estimates of the correlation function at radius $r_i$.  These
correlations would be required in order properly to quantify the
significance of any departure from gaussian statistics.  Note,
however, that the correlation matrix is very close to diagonal, so
distinct estimators are almost uncorrelated.
We have also tested the approximate correlation function considered in 
3D by \scite{BBKS} and in 2D by \scite{BE87}.  In this approximation,
more analytic progress is possible.  The approximation ignores
all derivatives of the temperature autocorrelation function, and it
should be accurate for large separations.  In the notation of the
appendix, this means setting all the $\lambda_{\alpha\beta\eta}=0$
except for the normalised autocorrelation function itself,
$\lambda_{000}$.  We show an example of the approximation, with the
accurate solution, in Fig. \ref{MDMNum}.   The approximation
does quite well, especially at large separations, but fails to
reproduce the oscillations.
\begin{figure}
\centering
\begin{picture}(200,200)
 \includegraphics{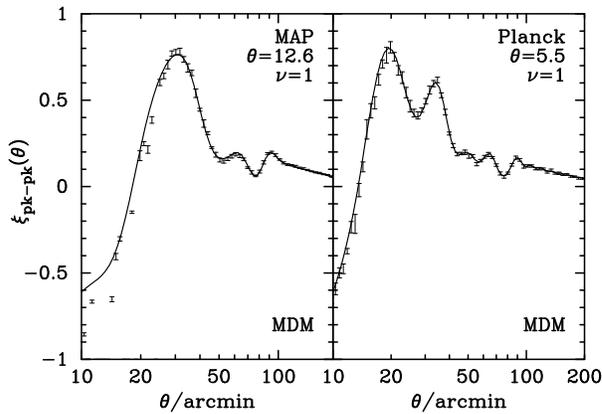}
\end{picture}
\caption[]{\label{MAPPlanck} Mixed dark matter model correlation
functions with errors expected from MAP (left) and Planck (right).
The MAP simulations have a 12.6 arcminute beam, and Planck 5.5
arcminutes.  Pixel sizes are taken to be 5 and 2 arcminutes.  The
simulations were done on square grids of side 3072 pixels, and the
error bars (from run-to-run variations) rescaled to mimic 75\% sky
coverage.  Note that the large pixels in MAP introduce discreteness
errors for separations up to about 20 arcminutes, and the theoretical MAP 
curve is reliable only for separations greater than 15 arcminutes.}  
\end{figure}
In Fig. \ref{MAPPlanck}, we show the expected errors from MAP and
Planck for a mixed dark matter model.  The data were obtained by
repeatedly simulating part of the sky on a grid, and rescaling the
errors to simulate 75\% sky coverage.  Details are given in the
caption.  

\section{Discussion}

In this paper we have shown that the peak-peak correlation function
for the microwave sky can be computed to high accuracy for gaussian
fields.  The only approximation we make is that the sky is flat, but
since the features in the correlation function appear on scales less
than about 2 degrees, the computations should be very accurate.  With
upcoming high-resolution CMB experiments MAP and Planck, these results
should allow a high-precision test of the hypothesis that structure
grows from gaussian initial fluctuations, such as predicted in most
inflationary models.  We have not computed the expected peak-peak
correlation function from defect models, but it is clear from the
abundance of structure in the correlation function that agreement with
these predictions, for all peak heights, would constitute a powerful
argument in favour of gaussian models.  We have also investigated
numerically the errors expected from MAP and Planck; many of the
features in the correlation function should be detectable with high
significance.  We note the similarity in the number of features in the
peak-peak correlation function as in the power spectrum (although
there may not be a simple correspondence between the two).  The use of
the two curves will, however, be rather different.  To fit the power
spectrum, one has the freedom to adjust fifteen or more parameters,
under the gaussian hypothesis.  With the peak-peak correlation
function, there are no free parameters; for a gaussian process, all
the statistics are specified once the power spectrum is fixed.  Thus
there is no freedom.  Note also that the information required for the
gaussian test is just the power spectrum; it does not rely on a
satisfactory parameter estimation.  For practical purposes, there are
some issues still to address; detector noise can by easily
incorporated into the analysis provided it is gaussian.  Practical
de-striping algorithms (e.g. \pcite{Delab98a}, \pcite{Revenu98}) to
reduce $1/f$ noise, or methods to remove sidelobe contamination
\cite{Delab98b} will need to be tested numerically to ensure that
residual non-gaussian contributions are small.  One interesting
possibility for dealing with residual point sources in the maps is to
put some constraint on the curvature at the peak (through the
parameter $x$) to exclude sharply-peaked sources.  The same trick
could be used to exclude very flat-topped peaks whose precise
locations may be difficult to determine accurately after dust
foreground subtraction.  Of course, both these techniques can be
applied to the data as well as incorporated into the theory.  There is
an alternative to this procedure for point sources uncorrelated with
the CMB; the peak distribution is then a superposition of two point
processes, and provided the flux and clustering properties of the
point sources are known, their contribution to the power spectrum can
be subtracted.  The gaussian test made on the joint distribution, as
the correlation function will be a suitably-weighted combination of
the two components.  An aside which is probably not relevant for this
application is that the analysis applies equally to any monotonic local
functions of 2D gaussian fields, as nothing changes except the
interpretation of the threshold, although in general the
non-gaussianity would be revealed in the one-point temperature
distribution.  In addition, one might expect any transformation
process to be local in 3D, in which case the mixing of information
through the last scattering surface would destroy the local nature.

\appendix

\section*{Appendix}

\subsection*{Joint distribution of variables}

Consider two points separated by $r$.  For convenience, place them on
the $x$-axis.   In the spirit of \scite{RS95}, we define
complex quantities 
\be
y_m^n(\bx) \equiv {i^n\over \sigma_n} \int d^2\bk \,\delta_\bk
\exp(i\bk\cdot \bx) \exp(im\theta) k^n
\ee
where $\theta$ is the angle between $\bk$ and the $x$-axis.  With this 
definition, $y_{-m}^n(\bx) = y_m^{n*}(\bx)$.  Then we require the following
complex variables:
\ba
\sigma_0 y_0^0 & = & \delta\nn
\sigma_1 y_1^1 & = & \delta_x+i\delta_y\nn
\sigma_2 y_0^2 & = & \delta_{xx}+\delta_{yy}\nn
\sigma_2 y_2^2 & = & \delta_{xx}+\delta_{yy}+2i\delta_{xy}
\ea
The correlations of these (at the same point) are
\be
\langle y_m^n(\bx)y_{m'}^{n'*}(\bx) \rangle =  {i^{n-n'}\over
\sigma_n\sigma_{n'}}\int d^2\bk d^2\bk' \langle \delta_\bk
\delta_{\bk'}\rangle \exp(i \bk\cdot \bx-i \bk'\cdot \bx)
\exp(im\theta -im'\theta') k^{n+n'}
\ee
With the definition of the power spectrum (\ref{Pk}), the integral
simplifies to 
\be
\langle y_m^n(\bx)y_{m'}^{n'*}(\bx) \rangle =  {i^{n-n'}\over
\sigma_n\sigma_{n'}}\sigma^2_{(n+n')/2}\delta^K_{mm'}
\ee
where $\delta^K$ is the Kronecker delta.  The cross-correlations
between variables at $\bx$ and at $\bx+\br$ are:
\be
\langle y_m^n(\bx)y_{m'}^{n'*}(\bx+\br) \rangle =  {i^{n-n'}\over
\sigma_n\sigma_{n'}}\int d^2\bk d^2\bk' \langle \delta_\bk
\delta_{\bk'}\rangle \exp(i \bk\cdot \bx-i \bk'\cdot (\bx+\br))
\exp(im\theta -im'\theta') k^{n+n'}.
\ee
Inserting the power spectrum removes the $\bk'$ integration, leaving a
2D integration over $\bk$.  The angular part of this is
\ba
I_\theta & \equiv &\int_0^{2\pi} d\theta
\exp(-ikr\cos\theta)\left[\cos(m-m')\theta + i\sin(m-m')\theta\right]\nn
&=&2\pi i^{|m-m'|}J_{|m-m'|}(-kr)
\ea
where $J$ is a Bessel function, and $J_\ell(-z) = (-1)^\ell
J^*_\ell(z)$.  Thus $\langle y_m^n(\bx+\br)y_{m'}^{n'*}(\bx) \rangle=
(-1)^{m+m'}\langle y_m^n(\bx)y_{m'}^{n'*}(\bx+\br) \rangle$ and we
find the cross-correlation matrix 
\be
\left(
\begin{array}{cccccc}
\lambda_{000}    &  -\lambda_{020}  &-\lambda_{011}    & -\lambda_{011}  &\lambda_{022}    & \lambda_{022} \\
-\lambda_{020} & \lambda_{220} &\lambda_{121} & \lambda_{121}& -\lambda_{222} & -\lambda_{222} \\
\lambda_{011}    &  -\lambda_{121}  &-\lambda_{110}    & -\lambda_{112}  &-\lambda_{121}    & \lambda_{123} \\
\lambda_{011}    &  -\lambda_{121}  &-\lambda_{112}    & \lambda_{110}  &\lambda_{123}    & -\lambda_{121} \\
\lambda_{022}    &  -\lambda_{222}  &-\lambda_{121}    & -\lambda_{123}  &\lambda_{220}    & \lambda_{224} \\
\lambda_{022}    &  -\lambda_{222}  &-\lambda_{222}    & \lambda_{121}  &\lambda_{224}    & \lambda_{220} \\
\end{array}
\right)
\ee
where the order of the variables is $\{y_0^0, y_0^2, y_1^1, y_1^{-1},
y_2^2, y_2^{-2}\}$ with quantities evaluated at $\bx$ down the left,
and at $\bx+\br$ along the top.
\be
\lambda_{\alpha\beta\eta} \equiv {(2\pi)^3\over
\sigma_\alpha\sigma_\beta} \int dk\, k^{1+\alpha+\beta} P(k) J_\eta(kr)
\ee
We now make real variables
\ba
\nu & = & y_0^0 \nn
X  & = & -y_0^2\nn
Y & = & {1\over 2}(y_2^2+y_{-2}^2)\nn
Z& = & {1\over 2i}(y_2^2-y_{-2}^2)\nn
\eta_x & \equiv & {1\over 2}(y_1^1+y_{-1}^1)\nn
\eta_y & = & {1\over 2i}(y_1^1-y_{-1}^1)\nn
\ea
These definitions agree with (\ref{Vars}) in the main text.  The last
two are $(\dd \delta/\dd x) /\sigma_1$ and $(\dd \delta/\dd y)
/\sigma_1$.  
The entire correlation matrix for the variables in the order
$(\nu_1,X_1,\eta_{x1},Y_1,\nu_2, X_2, \eta_{x2}, Y_2, \eta_{y1},
\eta_{y2}, Z_1, Z_2)$ splits into an $8\times 8$ matrix (top left) 
\be
M_8 \equiv \left(
\begin{array}{cccccccc}
1 & \gamma & . & . &\lambda_{000}    &  \lambda_{020}  &-\lambda_{011}& \lambda_{022} \\
\gamma & 1 & . & . &\lambda_{020} & \lambda_{220} &-\lambda_{121} &\lambda_{222} \\
 . & .&1/2&.&\lambda_{011}    &  \lambda_{121}&(\lambda_{110}-\lambda_{112})/2  &-(\lambda_{121}- \lambda_{123})/2
\\
 . & .&.&1/2&\lambda_{022}    &  \lambda_{222}  &(\lambda_{121}
-\lambda_{123})/2  &(\lambda_{220} +\lambda_{224})/2\\
 \lambda_{000}    &  \lambda_{020}  &\lambda_{011}& \lambda_{022} & 1 &
\gamma & . & . \\
\lambda_{020} & \lambda_{220} &\lambda_{121} &\lambda_{222} &  \gamma & 1&.\\
-\lambda_{011}    &  -\lambda_{121}&(\lambda_{110}-\lambda_{112})/2
&(\lambda_{121}- \lambda_{123})/2&.&.&1/2 & .\\
\lambda_{022}    &  \lambda_{222}  &-(\lambda_{121}
-\lambda_{123})/2  &(\lambda_{220} +\lambda_{224})/2 & .&.&.& 1/2\\
\end{array}
\right)
\ee
and a $4 \times 4$ block (bottom-right):
\be
M_4 \equiv \left(
 \begin{array}{cccc}
1/2&(\lambda_{110}+\lambda_{112})/2&.&-(\lambda_{121}+\lambda_{123})/2\\
(\lambda_{110}+\lambda_{112})/2&1/2&(\lambda_{121}+\lambda_{123})/2&.\\
. & (\lambda_{121}+\lambda_{123})/2&1/2&(\lambda_{220}-\lambda_{224})/2\\
-(\lambda_{121}+\lambda_{123})/2&.&(\lambda_{220}-\lambda_{224})/2&1/2\\
\end{array}
\right)
\ee
 Off-diagonal blocks are all zero.  The $4 \times 4$ matrix is readily
 inverted.  We used Mathematica to form the quadratic $v_i
 (M_8)^{-1}_{ij} v_j$ (where $\bv$ is the relevant part of the
 variable list, with the first derivatives set to zero).  Since it
 produces several thousand lines of Fortran, it is not reproduced
 here.

\bibliographystyle{mnras}
\bibliography{general}

\end{document}